# 盗聴者の視点からの量子鍵配送の安全性

岩越 丈尚†

†玉川大学 量子情報科学研究所 〒194-8610 東京都 町田市 玉川学園 6-1-1
E-mail: †t.iwakoshi@lab.tamagawa.ac.jp

**あらまし** 2005 年には量子鍵配送の安全性は，配送されるべき理想的な量子状態と実際に配送される量子状態との間のトレース距離で評価されるべきであると R. Renner らが提唱し，その際に提示された「トレース距離は量子鍵配送の最大失敗確率である」という解釈が広く受け入れられてきた．一方で 2009 年から，H. P. Yuen と O. Hirota らがそのような解釈は正しくないと主張し続けてきた．両者の主張を踏まえて，本研究の発表者は 2014 年 5 月の QIT30 より，量子鍵配送の安全性評価にトレース距離を用いることについて一貫して疑義を提示し，盗聴者が配送される鍵を推定する確率で安全性を評価することが重要とする Yuen の主張を取り上げてきた．しかしながらこれまで具体的に鍵の推定確率を説明することはできていなかった．今回は盗聴者が鍵を推定する確率の平均値を評価する方法を紹介し，Yuen が 2010 年から主張し続けてきた結果と一致したことを述べる．これにより従来から広く受け入れられてきた，トレース距離単体を量子鍵配送に失敗する確率とみなす解釈のどこに不備があったかを明らかにする．さらに，トレース距離が「識別不可能性」という量であるとする解釈についても問題点を解説する．

**キーワード** 量子鍵配送，安全性証明，トレース距離，量子信号検出理論，識別不可能性

## 1. はじめに

秘密裏にメッセージを交換するため，離れた 2 者がいかに秘密鍵を共有するかという課題は，暗号の歴史とともに続いてきた重要な課題である．現在では公開鍵暗号の登場によりこの課題に対する答えが提示されているが，公開鍵暗号は計算量的な問題であり，解読アルゴリズムが向上すればその安全性は脅かされる．

したがって，情報理論的に安全な鍵配送方法として量子鍵配送(QKD)[1]が注目を浴びている．当初は正規ユーザーと攻撃者との間の相互情報量を任意に小さくすることで QKD の無条件安全性を達成できるとされていたが[2]，攻撃者が量子メモリを保持している場合には配送された鍵を盗むことができることが明らかになった[3]．そこで[3]は，本来配送されるべき理想的な量子状態と実際に配送される量子状態の間のトレース距離を任意に小さくすることで本問題を克服できると提案した．さらにトレース距離には，配送された量子状態が理想的な量子状態とは異なる確率とみなせるという解釈が与えられ[3,4]，広く受け入れられてきた[5]．

その一方で，H. P. Yuen と O. Hirota らが，トレース距離にはそのような意味はないとする批判が 2009 年から提出され続けてきた[6-10]．彼らが提起した QKD の課題は多岐にわたるが，中でもトレース距離の解釈に対する疑義は大きな課題のひとつと言える．

QKD は実際に利用することが想定されている技術である．最近では，車載ネットワーク[11]や医療データベース[12]に利用することが検討されている．しかし，もし QKD の安全性証明に本質的な問題があるならば，普及してから社会インフラに与える影響は大きい．

そこで本研究の著者は 2014 年 5 月の QIT30 から，上記の批判を一貫して検討してきた[13-17]．今回は，「QKD の安全性は攻撃者が鍵を推定する確率で評価すべきである」と Yuen が繰り返し述べてきた主張が妥当であるとの結果を得たので報告する．

## 2. One-Time Pad の安全性

QKD の目的は，C. E. Shannon により情報理論的安全性が証明された One-Time Pad (OTP) を実現することにある．本節では，まず OTP の安全性の定義を述べる．

$X$ を平文，$K$ を共有鍵，$C$ を暗号文とする．正規の送信者 Alice は暗号文を $C = X \oplus K$ により生成し，正規の受信者 Bob は暗号文を $X = C \oplus K$ により復号する．このとき，$K$ の生起確率が一様独立ならば，

$$\begin{aligned} \Pr(X,C) &= \Pr(X|C)\Pr(C) \\ \Pr(X,C) &= \Pr(C)\Pr(C) \\ \therefore \Pr(X|C) &= \Pr(X) \end{aligned} \quad (1)$$

このことは，盗聴者 Eve が暗号文 $C$ を見ても，平文に関するヒントは一切得られないことを意味している．これが情報理論的に安全な暗号における，完全秘匿性である[18-20]．

## 3. QKD の安全性証明の定義と種類
### 3.1. ε-security の定義

QKD における ε-security の定義は次で与えられる．

$$\tfrac{1}{2}\mathrm{tr}|\rho_{ABE} - \tau_{AB} \otimes \sigma_E| \leq \varepsilon \quad (2)$$

ここで $\tau_{AB} \otimes \sigma_E$ は，送るべき望ましい量子状態であり，

$\rho_{ABE}$ は実際に配送される量子状態である．これらは，Alice と Bob が鍵 $k_A, k_B$ を得るとして，次の量子状態で与えられる．

$$\rho_{ABE} := \sum_{k_A, k_B} \Pr(k_A, k_B)|k_A, k_B\rangle\langle k_A, k_B| \otimes \rho_E(k_A, k_B) \quad (3)$$

$$\tau_{AB} \otimes \sigma_E := \sum_k 2^{-|K|}|k,k\rangle\langle k,k| \otimes \sigma_E \quad (4)$$

もし $\varepsilon$ がゼロならば必然的に $\rho_{ABE} = \tau_{AB} \otimes \sigma_E$ であるが，これは Eve が量子状態 $\sigma_E$ を保持していても Alice と Bob の鍵 $k$ に対するヒントは何も得られないことを意味する．問題は $\varepsilon$ がゼロでない場合である．

## 3.2. 2 種類の QKD の安全性証明

QKD の安全性を証明するには，まずトレース距離を三角不等式で2つに分解するところから始まる．

$$\begin{aligned}&\tfrac{1}{2}\text{tr}|\rho_{ABE} - \tau_{AB} \otimes \sigma_E|\\ &\leq \tfrac{1}{2}\text{tr}|\rho_{ABE} - \varsigma_{ABE}| + \tfrac{1}{2}\text{tr}|\varsigma_{ABE} - \tau_{AB} \otimes \sigma_E|\end{aligned} \quad (5)$$

ここで $\varsigma_{ABE}$ は鍵 $k_A$ と $k_B$ が一致した量子状態で，

$$\varsigma_{ABE} := \sum_{k_A, k_B} \Pr(k_A, k_B)|k_A, k_A\rangle\langle k_A, k_A| \otimes \rho_E(k_A, k_B) \quad (6)$$

これにより次の結果を得る．

$$\begin{aligned}&\tfrac{1}{2}\text{tr}|\rho_{ABE} - \varsigma_{ABE}| + \tfrac{1}{2}\text{tr}|\varsigma_{ABE} - \tau_{AB} \otimes \sigma_E|\\ &\leq \varepsilon_{cor} + \varepsilon_{sec} \leq \varepsilon\end{aligned} \quad (7)$$

ここで $\varepsilon_{cor}$ は鍵が一致しない確率の上界とされており，$\varepsilon_{sec}$ は得られた鍵が理想的でない確率の上界であるとされている．それらの上界がさらに $\varepsilon$ で与えられる．

今は $\varepsilon_{sec}$ の項に注目しよう．2 種類の安全性証明のうち1つ目は，トレース距離の下界を以って得られた鍵が理想的にならない確率であるとする．2つ目は，フィデリティでトレース距離の上界を与え，それを理想的な鍵を得られない確率とする．次の節ではこれらの証明を紹介し，その問題点を述べる．

## 4. QKD の安全性証明とその問題点
### 4.1. トレース距離の下界を失敗確率とする証明

トレース距離の下界を QKD の失敗確率とする解釈は[3-5, 21]などで広く受け入れられてきたが，具体的に証明を与えたのは C. Portmann と R. Renner の 2014 年の論文[22]である．以下にその証明を概説する．

トレース距離から統計距離を得るためにオペレータ $\Gamma_k$ のセットを考える．任意のオペレータについて成立する次の関係式を使うと，

$$\begin{aligned}&\tfrac{1}{2}\text{tr}|\varsigma_{ABE} - \tau_{AB} \otimes \sigma_E|\\ &= \tfrac{1}{2}\sum_k \Gamma_k \text{tr}|\varsigma_{ABE} - \tau_{AB} \otimes \sigma_E|\\ &\geq \tfrac{1}{2}\sum_k |\text{tr}[\Gamma_k(\varsigma_{ABE} - \tau_{AB} \otimes \sigma_E)]| = \tfrac{1}{2}\sum_k |P(k) - 2^{-|K|}|\end{aligned} \quad (8)$$

次に以下の結合確率を導入する．

$$R(k,k') = R'(k,k') + \frac{1}{\tfrac{1}{2}\sum_{k_E}|P(k_E) - U(k_E)|}S(k)T(k') \quad (9)$$

$$\begin{aligned}S(k) &= P(k) - R'(k,k)\\ T(k') &= U(k') - R'(k',k')\\ R'(k,k') &= \begin{cases}\min[P(k), U(k')] & \text{if } k = k'\\ 0 & \text{otherwise}\end{cases}\end{aligned} \quad (10)$$

ここで $U(k') = 2^{-|K|}$ である．この結合確率を用いると，

$$\tfrac{1}{2}\sum_k |P(k) - 2^{-|K|}| \geq 1 - \sum_k R(k,k) = \Pr[K \neq K'] \quad (11)$$

最後の項が，理想的な鍵を得られない確率とされる．

### 4.2. 下界を失敗確率とする解釈の問題点

[3-5, 21, 22]では，(11)の最後の項が「Alice と Bob が理想的な鍵を得ることに失敗する確率」であると解釈されている．例えば[3]では次の記述が見られる．

> "$\varepsilon$ security has an intuitive interpretation: with probability at least $1 - \varepsilon$, the key $S$ can be considered identical to a perfectly secure key $U$, i.e., $U$ is uniformly distributed and independent of the adversary's information. In other words, Definition 1 guarantees that the key $S$ is perfectly secure except with probability $\varepsilon$."

つまり，攻撃者がどのような情報を入手しようが，鍵の生成分布が一様独立分布 $U$ になる確率が $1 - \varepsilon$ であるとしている．この解釈は次の問題点をはらんでいる．

導入された結合確率を考えてみると，これは実際に配送された量子状態から得る鍵と，配送したいが実際には配送されていない量子状態から得る鍵との相関を意味する．そのような相関が現実にあり得るか大きな疑問が残る．K. Kato は[23]でこの問題を精査し，

$R(k,k') = P(k)U(k')$，2 つは独立とする確率分布以外に

ありえないと結論づけている．さらにこのことは次の結論を導く．

$$\tfrac{1}{2}\sum_k |P(k) - U(k)| < \Pr[K \neq K'] \quad (12)$$

このことはすでに Yuen が指摘している[7]．つまり(12)により，QKD の失敗確率とされた項はトレース距離の下界にすらならないのである．

### 4.3. トレース距離の上界を失敗確率とする証明

トレース距離の上界をフィデリティで設定し，それを理想的な量子状態を得られない確率とする証明は M.

Koashi により 2009 年に与えられた[24, 25]．その概要を以下に述べる．(7)の $\varepsilon_{\text{sec}}$ の項はフィデリティにより次のように抑えられる．

$$\frac{1}{2}\text{tr}|\varsigma_{\text{ABE}} - \tau_{\text{AB}} \otimes \sigma_{\text{E}}| \leq \sqrt{1 - F(\varsigma_{\text{ABE}}, \tau_{\text{AB}} \otimes \sigma_{\text{E}})^2} \\ \leq \sqrt{1 - F(\varsigma_{\text{ABE}}, |\psi\rangle\langle\psi| \otimes \sigma_{\text{E}})^2} \quad (13)$$

$$|\psi\rangle := \sum_k 2^{-|K|/2}|k,k\rangle \quad (14)$$

(13)ではいかなる量子操作でもフィデリティ $F$ が増加するという単調性を用いた．次に Eve の系をトレースアウトしたフィデリティを考える．

$$\sqrt{1 - F(\varsigma_{\text{ABE}}, |\psi\rangle\langle\psi| \otimes \sigma_{\text{E}})^2} \geq \sqrt{1 - F(\text{tr}_{\text{E}} \varsigma_{\text{ABE}}, |\psi\rangle\langle\psi|)^2} \\ = \sqrt{1 - \langle\psi|(\text{tr}_{\text{E}} \varsigma_{\text{ABE}})|\psi\rangle} \quad (15)$$

(13)と比べると，(15)は逆向きの不等号を持っているが[16, 17]，[24, 25]は等号を実現するような $\sigma_{\text{E}}$ を選ぶことを説いている．次に $\text{tr}_{\text{E}} \varsigma_{\text{ABE}}$ に対してエンタングルメント抽出を行い，(14)の最大エンタングル状態に近い状態を得るのに失敗する確率を $\varepsilon_{\text{sec}}^2$ とすると，

$$\sqrt{1 - \langle\psi|(\text{tr}_{\text{E}} \varsigma_{\text{ABE}})|\psi\rangle} \leq \sqrt{1 - (1 - \varepsilon_{\text{sec}}^2)} = \varepsilon_{\text{sec}} \quad (16)$$

これにより[24, 25]は(7)が得られると結論づけた．

### 4.4. 上界を失敗確率とする解釈の問題点

4.1 節で説明した証明と比べると，不自然な仮定は本証明にはなさそうである．ところが[22]は，QKD を並行に走らせた場合に全体のシステムとしての安全性が，$\sigma_{\text{E}}$ が以下の場合に限り得られているとしている．

$$\sigma_{\text{E}} := \sum_{k_A,k_B} \Pr(k_A, k_B) \rho_{\text{E}}(k_A, k_B) \quad (17)$$

実際 M. Tomamichel は $\varepsilon$-security の定義[26]を(18)から(19)に変えたと本研究の著者に個人的に伝えた[27]．

$$\frac{1}{2}\min_{\sigma_{\text{E}}} \text{tr}|\rho_{\text{ABE}} - \tau_{\text{AB}} \otimes \sigma_{\text{E}}| \leq \varepsilon_{\text{sec}} \quad (18)$$

$$\frac{1}{2}\text{tr}|\rho_{\text{ABE}} - \tau_{\text{AB}} \otimes \sigma_{\text{E}}| \leq \varepsilon_{\text{sec}} \quad (19)$$

このことは $\sigma_{\text{E}}$ を自由に選べないことを意味している．すると(15)で等号を保持することができなくなり，$\varepsilon_{\text{sec}}$ がトレース距離の上界にならなくなるのである．

しかし(15)の等号が保持できるように $\sigma_{\text{E}}$ を選べたとしよう．これはどちらにせよトレース距離が非ゼロであることを意味しており必然的に $\varsigma_{\text{ABE}} \neq \tau_{\text{AB}} \otimes \sigma_{\text{E}}$ となる．このことは，配送された量子状態 $\varsigma_{\text{ABE}}$ が，鍵の一様独立分布を保証しないか，Eve の量子系がまだ鍵との相関を保持しているか，あるいはその両方を意味する．

### 4.5. トレース距離を識別不可能性とする解釈

この他にもトレース距離を「識別不可能性」という量であるとする解釈もある[28, 22]．本解釈は C. W. Helstrom による 2 元信号検出理論[29]を根拠とする．その概要を以下に示す．

Helstrom の理論によると，Alice は 2 つの量子状態 $\rho_0$ か $\rho_1$ を先験確率 $p_0$ と $p_1$ で準備し，Bob は受信基底を最適にすることでどちらが配送されたのかを判断する．このとき Bob が量子状態の区別に成功する確率は，

$$P_{\text{guess}} = \frac{1}{2} + \frac{1}{2}\text{tr}|p_0\rho_0 - p_1\rho_1| \quad (20)$$

ここで先験確率を $p_0 = p_1 = 1/2$ とすると

$$P_{\text{guess}} = \frac{1}{2} + \frac{1}{2} \times \frac{1}{2}\text{tr}|\rho_0 - \rho_1| \quad (21)$$

[28, 22]は以上から，トレース距離を下記のように解釈する．Alice は 2 つの QKD 装置を準備する．一方は Eve が量子状態と相互作用できるが，もう一方は Eve が相互作用できず，相互作用ができていると思わせるような出力を Eve に与えるインターフェースを有する．Alice は上記 2 つの装置を確率 1/2 で準備し，Eve はどちらの装置が使用されているのかをその出力から判断する．(21)において $\rho_0 = \varsigma_{\text{ABE}}$，$\rho_1 = \tau_{\text{AB}} \otimes \sigma_{\text{E}}$ とおくと，Eve が 2 つの量子状態を区別できる最大確率は(21)で与えられ，そのトレース距離の値が，当て推量で装置を判別する確率 1/2 からどれだけアドバンテージを得られるかという量，すなわち識別不可能性である．

### 4.6. 識別不可能性解釈の問題点

識別不可能性の問題点は以下の通りである．Eve が 2 つの装置のどちらが使用されたかを推定する確率は(21)の $P_{\text{guess}}$ であるが，正しく推定したところで Eve が Alice と Bob の鍵についてどう推定するのか議論を進められない．さらに次の問題点をはらんでいる．

QKD が運用される状況に立ち返って考えてみよう．第一に，$\tau_{\text{AB}} \otimes \sigma_{\text{E}}$ は配送したい量子状態ではあるが実際には配送されない量子状態であり，実際に配送される量子状態が $\varsigma_{\text{ABE}}$ である．しかし識別不可能性の文脈ならば，これらの量子状態を先験確率 1/2 で配送しなければならない．この点が識別不可能性の文脈と QKD の文脈とで合致しない．さらにもし Eve が相互作用できない QKD 装置を確率 1/2 で準備できるならば，そのような装置をなぜ確率 1 で使用しないのかが不明である．一方で，実際の QKD では確率 1 で Eve が相互作用できるものを用いることになる．つまり，本解釈では，先験確率の付け方が恣意的である．第三に，2 元量子信号検出理論の文脈では Bob は配送される量子状態 $\rho_0$ または $\rho_1$ の全体を受信するのだが，QKD の文脈では Eve が保持するのは，例えば $\text{tr}_{\text{AB}} \varsigma_{\text{ABE}}$ などの部分系であり，この点で Helstrom の 2 元信号検出理論の物理学的文脈は QKD の文脈とは一致しない．

以上からトレース距離を識別不可能性とする解釈は，QKD の安全性を評価する上で有用とは言いがたい．

## 5. 盗聴者の視点からの量子鍵配送の安全性
### 5.1. 盗聴者による鍵の推定確率

[22]は下記(22)も導出しているが，その重要性については述べていない．

$$\Pr(K|E) \leq 2^{-|K|} + \tfrac{1}{2}\mathrm{tr}|\varsigma_{ABE} - \tau_{AB} \otimes \sigma_E| \leq 2^{-|K|} + \varepsilon_{\mathrm{sec}} \quad (22)$$

この確率は配送された鍵を Eve が正しく推定する確率の期待値である．導出方法は，下記のとおりである．トレース距離は一般に(23)の不等号を満たす．

$$\mathrm{tr}\left[\Gamma(\varsigma_{ABE} - \tau_{AB} \otimes \sigma_E)\right] \leq \tfrac{1}{2}\mathrm{tr}|\varsigma_{ABE} - \tau_{AB} \otimes \sigma_E| \quad (23)$$

ここで次のオペレータを導入する．

$$\Gamma = \sum_{k_A} |k_A, k_A\rangle\langle k_A, k_A| \otimes M_{k_A} \quad (24)$$

したがって，

$$\begin{aligned}
&\mathrm{tr}\left[\Gamma(\varsigma_{ABE} - \tau_{AB} \otimes \sigma_E)\right] \\
&= \sum_{k_A, k_B}\left[p(k_A, k_B)\Pr(k_A|E)\right] - 2^{-|K|} \\
&= \Pr(K|E) - 2^{-|K|}
\end{aligned} \quad (25)$$

$2^{-|K|}$ を移項すれば (22)が得られる．ここで気をつけて頂きたいのは，(22)の Eve の鍵の推定確率は，特定のプロトコルや実装，証明手順に一切依存しないことである．トレース距離により QKD の安全性を評価するかぎり，全ての安全性理論に当てはまる．

以上の結果は，次のことを明らかにしている．まず，これまでの「トレース距離は QKD の最大失敗確率」とする解釈では，確率 $1 - \varepsilon_{\mathrm{sec}}$ で理想的な確率分布を持った鍵が得られるとしていた．ところが Eve による鍵の推定確率は，$\varepsilon_{\mathrm{sec}} = 0$ でない限り一様独立分布 $2^{-|K|}$ にはならないのである．次に，Eve による鍵の推定確率は，$2^{-|K|}$ の項だけ明らかに「最大失敗確率」より大きくなりうる．そして最後に，この Eve による鍵の推定確率は Yuen がすでに指摘している結果[7]と一致する．

### 5.2. Eve による鍵の推定確率の定量的意味

次に，(22)が QKD の安全性について述べていることをより具体的に言及する．

[22]は「宇宙の年齢ほども続く安全性」として $\varepsilon/|K| = 10^{-24}$ を例示している．また，最近の実験結果[30]は，トレース距離 $2^{-50} = 8.9 \times 10^{-16}$ を達成したとしている．

ここで，QKD による典型的な秘密鍵の長さを $|K| = 10^6$ bits とする[26]．先の[22]の例示なら $\varepsilon_{\mathrm{sec}} \sim 10^{-18}$ となるが，これは次のことを意味する．

$$2^{-|K|} = 2^{-10^6} = 10^{-326228} \ll \varepsilon_{\mathrm{sec}} = 10^{-18} \quad (26)$$

前述したように，「トレース距離は QKD における最大失敗確率」とする解釈では，確率 $1 - \varepsilon_{\mathrm{sec}}$ で理想的な鍵の確率分布 $2^{-|K|}$ が得られるとしていた．しかし Eve による鍵の推定確率(22)と，その数値例(26)の意味は，$\max \Pr(K|E) = \varepsilon_{\mathrm{sec}} + 2^{-|K|}$ は一様独立分布 $2^{-|K|}$ から程遠いということであり，$\varepsilon_{\mathrm{sec}} = 0$ のときのみ一様独立分布になるということである．したがって，2 節で説明した完全秘匿性の条件 $\Pr(K|E) = \Pr(K)$ を QKD で配送された鍵は満たさないということになる．このことは Universal Composability [3-5, 21, 22, 28]という概念にも影響する．本概念では，配送された鍵が一様独立であるため，その一部を他の暗号システムやプロトコルで使っても安全性には影響がないとするが，上述の通り，満たすべき一様独立性は Eve から見れば $\varepsilon_{\mathrm{sec}} = 0$ でないかぎり達成されない．

もちろん，鍵が完全でなくても Eve による推定確率が十分小さければ安全性には問題がない，とする考えもありうる．この考え方は適切といえるが，具体的にどのくらい推定確率が小さければ十分と言えるのかを $\varepsilon_{\mathrm{sec}} = 2^{-50} = 8.9 \times 10^{-16}$ [30]という値を用いて例示する．

QKD の装置が 1 年間稼働し続けたとして，その通信速度が $10^9$ bits/sec，秘密鍵の長さが $10^6$ bits であるとすると，1 年に $3 \times 10^{10}$ 個の鍵が配送されることになる．先の $\varepsilon_{\mathrm{sec}}$ の値と $2^{-|K|}$ の値から計算すると，年間の盗聴件数の期待値は $3 \times 10^{-5}$ 件となる．この値は十分小さいように見える．

一方で，2008 年に日本で起きた交通事故による死亡数は $7.5 \times 10^3$ 件であり[31]，この年の登録自動車台数は $7.9 \times 10^7$ 台である[32]．すると 1 年間に 1 台の自動車が起こす死亡事故数は，$9.5 \times 10^{-5}$ 件であったことになる．

以上の概算から，$\varepsilon_{\mathrm{sec}} = 2^{-50} = 8.9 \times 10^{-16}$ であったとしても，QKD の装置が年間に被りうる盗聴件数と自動車による年間死亡事故件数はだいたい同じである．もちろん，秘密鍵の長さ $10^6$ bits に対して QKD の装置の通信速度が $10^9$ bits/sec よりも低ければ，必ずしも同じではないが，QKD の実用化を目指しているのであれば，概算値が大きく違うこともないであろう．冒頭で述べたように QKD が自動車並みに世界に普及する未来になれば，自動車と同程度の事故も発生するのである．

$\varepsilon_{\mathrm{sec}} = 2^{-50}$ を達成してようやく上記のような安全性であるので，社会の重要なインフラとして使われるにはまだ QKD の安全性が十分ではないと思われる．

### 5.3. 安全性に関するその他の課題

他にも Yuen はさまざまな重要な指摘を行っている．例えば，[33]では古典回線上で交換される情報 $y$ で平均されたトレース距離で安全性を評価しているが，

$$\sum_y P_{\text{pub}}(y) \tfrac{1}{2} \text{tr} \left| \rho_{\text{ABE}|y} - \tau_{\text{AB}|y} \otimes \sigma_{\text{E}|y} \right|$$
$$\leq \sum_y P_{\text{pub}}(y)\sqrt{2P_{\text{ph}|y}} \leq \sqrt{\sum_y P_{\text{pub}}(y) 2P_{\text{ph}|y}} = \sqrt{2P_{\text{ph}}} := \varepsilon \qquad (27)$$

Yuenは理論上の$\varepsilon$が達成されても実際には$3\varepsilon^{1/3} \gg \varepsilon$程度の安全性しかないとしている[9]．これは次の理由による．例えば$y$がプライバシー増幅コードであるとしよう．このときEveは$y$の平均値を見ているわけではなく，交換された$y$がなんであるかを具体的に知っている．つまり平均されたトレース距離でなく，平均される前のトレース距離で安全性を評価しなくてはならない．Yuenはこの問題に対して，複数回のMarkov不等式を適用すれば安全性を評価できるとしている．プライバシー増幅コードで1回の平均化がなされるなら，平均される前のトレース距離は少なく見積もっても$2\varepsilon^{1/2}$，誤り訂正コードでさらに平均化されているのであれば，あるいは後述の既知平文攻撃での平均化を考慮するならば，さらに$3\varepsilon^{1/3}$となる．

他の問題として，Yuenは前述の既知平文攻撃[18-20]を挙げる．既知平文攻撃とは，暗号化される前の平文の一部を知っている状況下で，残りの平文の安全性を評価するという課題である．実際，第二次世界大戦中では，パープル暗号と呼ばれた旧日本軍の暗号は解読不能とされながら，既知平文攻撃で破られた[34]．より身近な例ではネットワークを流れるパケットのヘッダなどはほぼ定型文書であり既知平文攻撃の対象となる．もともと[3]は，攻撃者との相互情報量が小さくても，攻撃者が量子メモリを持っている場合には，既知平文攻撃で残りの鍵を入手できることを示しており，既知平文攻撃の可能性は考慮しなくてはならない．従来の「トレース距離はQKDの最大失敗確率」とする解釈では，確率$1-\varepsilon_{\text{sec}}$で鍵の生成確率分布は一様独立であるため，既知平文攻撃で鍵の一部が知られても残りの鍵の安全性は変わらないとされた[22]．しかしEveによる鍵の推定確率は$\varepsilon_{\text{sec}} = 0$でない限り一様独立でないため，上記のような単純な問題ではなくなる．この課題に対してYuenはすでに答えを導いている．

さらにYuenは次の状況も課題として挙げている．例えばEveが推定した鍵は実際の秘密鍵とは異なるが，それに近いものを推定したとする．Eveが推定した鍵を使ってOTPを復号すると，Eveにとっていくらかのそれに近いものを推定したとする．Eveが推定した鍵を使ってOTPを復号すると，Eveにとっていくらかのビット誤り率を含んだ平文に見える．このように，Eveにとってのビット誤り率がどこまでなら平文の安全性が許容できるかというのが，Yuenの呼ぶところの"Bit-Error-Rate issue"である[8]．当然，いくらかの誤りを含んだ鍵までが可能性に考慮されるため，(22)で計算されたEveによる鍵の推定確率より大きな確率を考慮することになる．これにどのように上界を与えるかという課題には，Yuenも答えを出していない．

Yuenは上述した他に多くの課題がQKDには残されているとする．ここで挙げた項目以外については，Yuenの他の論文を含めて参照されたい．

## 6. まとめ

量子鍵配送におけるトレース距離は，理想的な鍵を得ることに失敗する確率を与えると解釈されてきた．

しかしながら，そのような解釈は正しくないという批判がH. P. YuenとO. Hirotaにより2009年から繰り返し提出されてきた．本研究の著者は，上記の批判の検証を2014年5月のQIT30から行ってきたが，Yuenが繰り返し重要性を主張してきたEveによる鍵の推定確率が今回の研究で明らかになったため，YuenとHirotaの批判が正しいと結論付ける．

一方で，トレース距離を識別不可能性とする解釈もあるが，この解釈は量子鍵配送の安全性を評価する上で有用な測度とは言いがたいことも説明した．

本研究の著者はこうした知見を量子鍵配送の安全性を評価する上で真剣に議論して頂けることを望んでいる．

## 文　献

[1] C. H. Bennett, G. Brassard, "Quantum cryptography: Public key distribution and coin tossing," Proceedings of IEEE International Conference on Computers, Systems and Signal Processing, Vol. 175. No. 0. (1984), or Theoretical Computer Science 560 7-11 (2014).

[2] C. H. Bennett, et al. "Generalized privacy amplification," Information Theory, IEEE Transactions on 41.6 1915-1923 (1995).

[3] R. König, et al. "Small accessible quantum information does not imply security." Physical Review Letters 98.14 140502 (2007).

[4] R. Renner, R. König, "Universally composable privacy amplification against quantum adversaries," Theory of Cryptography. Springer Berlin Heidelberg, 407-425 (2005).

[5] V. Scarani, et al. "The security of practical quantum key distribution," Reviews of modern physics 81.3 1301 (2009).

[6] H. P. Yuen, "Universality and The Criterion 'd' in Quantum Key Generation," arXiv:0907.4694v1, quant-ph (2009).

[7] H. P. Yuen, "Fundamental Quantitative Security In Quantum Key Distribution," Physical Review A 82.6 062304 (2010).

[8] H. P. Yuen, "What The Trace Distance Security Criterion in Quantum Key Distribution Does And Does Not Guarantee," arXiv:1410.6945v1, (2014).

[9] H. P. Yuen, "Problems of Security Proofs and Fundamental Limit on Key Generation Rate in Quantum Key Distribution," arXiv:1205.3820v2, quant-ph (2012).

[10] O. Hirota, "Incompleteness and Limit of Quantum Key Distribution Theory," arXiv:1208.2106v2, quant-ph (2012).

[11] Updating Quantum Cryptography and Communications2015 in QCrypt2015 (2015).

[12] 東芝プレスリリース，"盗聴が理論上不可能な量子暗号通信システムの実証試験の開始について，" http://www.toshiba.co.jp/about/press/2015_06/pr_j1801.htm (2015).

[13] 岩越 丈尚，広田 修，"トレース距離を量子鍵配送の失敗確率とする解釈における課題，" 第 30 回 量子情報技術研究会，電子情報通信学会 IEICE QIT2014-9 (2014).

[14] 岩越 丈尚，広田 修，"トレース距離を量子鍵配送の失敗確率とする解釈における課題: Part II," 電子情報通信学会 IEICE ISEC2014-48 (2014).

[15] T. Iwakoshi, O. Hirota, "Misinterpretation of statistical distance in security of quantum key distribution shown by simulation," SPIE Security+ Defence. International Society for Optics and Photonics, pp. 92540L-92540L (2014).

[16] 岩越 丈尚，"相補性を用いた量子鍵配送の安全性証明の欠陥 -トレース距離を量子鍵配送の失敗確率とする解釈における課題: Part III-," 第 31 回 量子情報技術研究会，電子情報通信学会 IEICE QIT2014-58 (2014).

[17] 岩越 丈尚，"相補性を用いた量子鍵配送の安全性証明の欠陥: Part II," 第 32 回 量子情報技術研究会，電子情報通信学会 IEICE QIT2015-16 (2015).

[18] D. R. Stinson, 櫻井 幸一 監訳, "暗号理論の基礎," 共立出版株式会社，初版第二刷 (1997).

[19] A. J. Menezes, P. C. van Oorschot, and S. A. Vanstone, "Handbook of applied cryptography," CRC Press (1997).

[20] J. A. Buchmann, 林 芳樹 訳, "暗号理論入門," 原書第 3 版, Springer Japan, (2008).

[21] R. Renner, "Security of Quantum Key Distribution," International Journal of Quantum Information 6.01, 1-127 (2008).

[22] C. Portmann, R. Renner, "Cryptographic security of quantum key distribution," arXiv:1409.3525v1, quant-ph (2014).

[23] K. Kato, "Coupling Lemma and Its Application to The Security Analysis of Quantum Key Distribution," arXiv:1505.06269v1, quant-ph (2015), or Tamagawa University Quantum ICT Research Institute Bulletin, vol.4, no.1, pp.23-30, (2015).

[24] M. Koashi, "Simple security proof of quantum key distribution based on complementarity," New Journal of Physics 11.4 045018 (2009).

[25] 小芦 雅斗，小柴 健史, "量子暗号理論の展開," サイエンス社 (2008).

[26] M. Tomamichel, et al., "Tight finite-key analysis for quantum cryptography," Nature communications 3 634 (2012).

[27] M. Tomamichel, et al., "Tight finite-key analysis for quantum cryptography," arXiv:1103.4130v2, quant-ph (2012).

[28] M. Ben-Or, et al. "The universal composable security of quantum key distribution." Theory of Cryptography, Springer Berlin Heidelberg, p.386-406. (2005).

[29] C. W. Helstrom, "Quantum Detection and Estimation Theory," Journal of Statistical Physics 1.2 231-252 (1969) or Academic press (1976).

[30] H. Takesue, et al., "Experimental quantum key distribution without monitoring signal disturbance," Nature Photonics (2015).

[31] 厚生労働省による調査, http://www.mhlw.go.jp/toukei/saikin/hw/jinkou/tokusyu/furyo10/01.html

[32] 一般財団法人自動車検査登録情報協会の調査, https://www.airia.or.jp/publish/file/e49tph00000004sb-att/e49tph00000004si.pdf

[33] M. Hayashi, T. Tsurumaru, "Concise and tight security analysis of the Bennett–Brassard 1984 protocol with finite key lengths," New Journal of Physics 14.9 093014 (2012).

[34] 吹田 智章, "暗号のすべてがわかる本," 株式会社技術評論社 (1998)

# Security of Quantum Key Distribution from Attacker's View

Takehisa IWAKOSHI†

†Quantum ICT Research Institute, Tamagawa Univ.　6-1-1 Tamagawa-Gakuen, Machida-Shi, Tokyo, 194-8610 Japan
E-mail:　†t.iwakoshi@lab.tamagawa.ac.jp

**Abstract**　In 2005, trace distance between an ideal quantum state to be distributed and an actually distributed quantum state was introduced as a valid security measure of Quantum Key Distribution (QKD) by R. Renner et al., then it has been perceived that the trace can be interpreted as a maximum failure probability of QKD. While such a perspective has been widely accepted, H. P. Yuen and O. Hirota have been warning that such an interpretation is not correct since 2009. The author of this study has been giving questions on the interpretation of the trace distance based on their criticisms since QIT30 in May 2014, and has been proposing Yuen's idea to evaluate the security of QKD by the probability for the attacker to guess the correct key. However, the author could not give the guessing probability concretely. In this study, the author explains how to derive the average guessing probability for the attacker, where its result equals to Yuen's derivation firstly seen in 2010. From this result, one will see the problems with the maximum failure probability interpretation of the trace distance clearly. This study also explains the indistinguishability advantage interpretation is also invalid.

**Keyword**　Quantum Key Distribution, Security Proof, Trace Distance, Quantum Detection Theory, Indistinguishability

## 1. Introduction

Since the beginning of the history of cryptography, how the distant parties communicate secretly has been a serious issue. Nowadays, public-key encryptions solved the issue. However, public-key cryptographies rely on the computational complexity, therefore improvements in crypto-analysis algorithms necessarily threatens their security.

As a consequence, Quantum Key Distribution (QKD) [1] has been attracting attentions since it would offer information-theoretic security. At first, the security of QKD protocols were guaranteed by arbitrarily small mutual information between the legitimate users and the eavesdropper [2], however, it was proven that the eavesdropper could obtain the distributed key utilizing quantum memory [3]. Therefore the study [3] also showed that giving upper-bound to trace distance between the provided quantum states and an ideal quantum states expected to be provided with a negligibly small parameter would solve the problem. Ref. [3] and Ref. [4] also gave an interpretation that the upper-bound of the trace distance could be interpreted as a maximum failure probability where the distributed quantum state would not be identical to the ideal quantum state to be distributed, then the interpretation has been widely accepted [5].

However, H. P. Yuen and O. Hirota have been claiming since 2009 that the above interpretation is incorrect [6-10]. Their claim consists on many parts, but the most important part is that the trace distance does not give such an operational interpretation.

QKD is expected to be used in the real world. Recently researchers are aiming to apply QKD systems to onboard communication systems [11] and medical data centers [12]. However, if there is a imperfections in the security proofs of QKD, the impact on these infrastructures will be very sever.

Therefore, the author of this study has started analyses of these criticisms since the QIT30 conference in May 2014 [13-17]. This time, the author reports that the criticisms are correct and shows that the claim given by Yuen is valid that the security of QKD should be evaluated by the probability for the eavesdropper to guess the secret key.

## 2. Security of One-Time Pad

The goal of QKD is to obtain a secret key for One-Time Pad (OTP), which gives perfect secrecy as shown by C. E. Shannon. Therefore, this section shows briefly how the security of OTP is defined.

Let $C$ be a ciphertext, let $X$ be a plaintext, and let $K$ be the shared secret key. Then, the legitimate transmitter Alice generates the ciphertext by $C = X \oplus K$, while the legitimate receiver Bob obtains the plaintext by $X = C \oplus K$. If the probability distribution of a bit string $K$ is independent and identically distributed (IID), the following condition is satisfied.

$$\Pr(X, C) = \Pr(X|C)\Pr(C)$$
$$\Pr(X, C) = \Pr(X)\Pr(C) \quad (1)$$
$$\therefore \Pr(X|C) = \Pr(X)$$

This means that the eavesdropper who obtained $C$ has no hints in guessing possible plaintexts $X$. This is the perfect secrecy of information-theoretical secure cryptographies [18-20].

## 3. Interpretations of the security of QKD
### 3.1. Definition of $\varepsilon$-security
$\varepsilon$-security of QKD is defined as follows.

$$\tfrac{1}{2}\mathrm{tr}|\rho_{ABE} - \tau_{AB} \otimes \sigma_E| \leq \varepsilon \quad (2)$$

Here, $\tau_{AB} \otimes \sigma_E$, is a desirable quantum state to be provided and $\rho_{ABE}$ is the actually distributed quantum state. More concretely, when Alice and Bob obtain the final key $k_A$, $k_B$,

$$\rho_{ABE} := \sum_{k_A, k_B \in K} \Pr(k_A, k_B)|k_A, k_B\rangle\langle k_A, k_B| \otimes \rho_E(k_A, k_B) \quad (3)$$

$$\tau_{AB} \otimes \sigma_E := \sum_{k \in K} 2^{-|K|} |k,k\rangle\langle k,k| \otimes \sigma_E \quad (4)$$

If the $\varepsilon$ is zero, we necessarily obtain $\rho_{ABE} = \tau_{AB} \otimes \sigma_E$, which means Eve has no hints on the distributed key. At the same time, it confirms that the distributed key is IID, which confirms the security of OTP. However, problems arise when $\varepsilon$ is non-zero.

### 3.2. Types of security proofs of QKD
There are major two types of security proofs. Consider a triangle inequality as follows.

$$\tfrac{1}{2}\mathrm{tr}|\rho_{ABE} - \tau_{AB} \otimes \sigma_E| \\ \leq \tfrac{1}{2}\mathrm{tr}|\rho_{ABE} - \varsigma_{ABE}| + \tfrac{1}{2}\mathrm{tr}|\varsigma_{ABE} - \tau_{AB} \otimes \sigma_E| \quad (5)$$

Here, $\varsigma_{ABE}$ is a quantum state Alice's and Bob's key agree,

$$\varsigma_{ABE} := \sum_{k_A, k_B \in K} \Pr(k_A, k_B)|k_A, k_A\rangle\langle k_A, k_A| \otimes \rho_E(k_A, k_B) \quad (6)$$

Therefore we obtain

$$\tfrac{1}{2}\mathrm{tr}|\rho_{ABE} - \tau_{AB} \otimes \sigma_E| \\ \leq \Pr[K_A \neq K_B] + \tfrac{1}{2}\mathrm{tr}|\varsigma_{ABE} - \tau_{AB} \otimes \sigma_E| \quad (7) \\ \leq \varepsilon_{\mathrm{cor}} + \varepsilon_{\mathrm{sec}} \leq \varepsilon$$

Here, $\varepsilon_{\mathrm{cor}}$ is regarded as the upper-bound of the probability where Alice's and Bob's key do not meet, $\varepsilon_{\mathrm{sec}}$ is regarded as the upper-bound of the probability where the obtained final key is not desirable. Furthermore, $\varepsilon_{\mathrm{cor}}$ and $\varepsilon_{\mathrm{sec}}$ are upper-bounded by $\varepsilon$.

Now we focus on the $\varepsilon_{\mathrm{sec}}$ term. There are two types of security proofs. The first type regards the lower-bound of $\varepsilon_{\mathrm{sec}}$ as the probability where Alice and Bob cannot obtain a desirable key. The second type regards the upper-bound of $\varepsilon_{\mathrm{sec}}$ as the probability where Alice and Bob cannot obtain a desirable key by converting the trace distance into fidelity. The next section will show the concrete steps of the proofs and reasons why these interpretations have problems.

## 4. Security Proofs of QKD and Problems
### 4.1. The Lower-Bound of Trace Distance being Interpreted as Failure Probability of QKD

The interpretation was given in Refs. [3-5, 21], then widely accepted among researchers. However, the first concrete proof was given in 2014 by C. Portmann and R. Renner in Ref. [22]. The following is the overview.

To obtain statistical distance from the trace distance, introduce a set of operators $\Gamma_k$. Arbitral operators satisfy the following inequality,

$$\tfrac{1}{2}\mathrm{tr}|\varsigma_{ABE} - \tau_{AB} \otimes \sigma_E| \\ = \tfrac{1}{2}\sum_k \Gamma_k \mathrm{tr}|\varsigma_{ABE} - \tau_{AB} \otimes \sigma_E| \quad (8) \\ \geq \tfrac{1}{2}\sum_k |\mathrm{tr}[\Gamma_k(\varsigma_{ABE} - \tau_{AB} \otimes \sigma_E)]| = \tfrac{1}{2}\sum_k |P(k) - 2^{-|K|}|$$

Then, introduce a following joint distribution.

$$R(k, k') = R'(k, k') + \frac{1}{\tfrac{1}{2}\mathrm{tr}|\varsigma_{ABE} - \tau_{AB} \otimes \sigma_E|} S(k)T(k') \quad (9)$$

$$S(k) = P(k) - R'(k, k') \\ T(k) = U(k) - R'(k, k') \quad (10) \\ S(k, k') = \begin{cases} \min[P(k), U(k')] & \text{if } k = k' \\ 0 & \text{otherwise} \end{cases}$$

Here, $U(k') = 2^{-|K|}$. From the joint distribution,

$$\tfrac{1}{2}\sum_k |P(k) - 2^{-|K|}| \geq 1 - \sum_k R(k, k) = \Pr[K \neq K'] \quad (11)$$

The last term is interpreted as the failure probability.

### 4.2. Problems with Interpretation of Lower-Bound of Trace Distance

According to [3-5, 21, 22], the last term in (11) was interpreted as "the probability where Alice and Bob fail in obtaining the ideal final key". For instance, the following explanation is seen in Ref. [3].

"$\varepsilon$ security has an intuitive interpretation: with probability at least $1 - \varepsilon$, the key $S$ can be considered identical to a perfectly secure key U, i.e., U is uniformly distributed and independent of the adversary's information. In other words, Definition 1 guarantees that the key S is perfectly secure except with probability ε."

However, this interpretation has the following problems.

The introduced joint probability has the correlation between the actually distributed quantum state and the ideal quantum state which is not actually distributed. It is questionable that there could be such a correlation in the real world. K. Kato precisely discussed about this problem [23] and concluded that only $R(\bm{k},\bm{k}') = P(\bm{k})U(\bm{k}')$ is acceptable, which gives

$$\tfrac{1}{2}\sum_{\bm{k}}|P(\bm{k})-U(\bm{k})| < \Pr[\bm{K} \neq \bm{K}'] \tag{12}$$

This is what Yuen already pointed out [7]. This means that the right hand side of (12) cannot be even the lower-bound of the trace distance.

### 4.3. The Upper-Bound of Trace Distance being Interpreted as Failure Probability of QKD

The interpretation that the upper-bound of the trace distance directly gives the maximum failure probability of QKD by converting it to fidelity was given by M. Koashi in 2009 [24, 25]. The following is the overview. The $\varepsilon_{\text{sec}}$ term in (7) has the following relation.

$$\tfrac{1}{2}\text{tr}|\varsigma_{\text{ABE}} - \tau_{\text{AB}} \otimes \sigma_{\text{E}}| \leq \sqrt{1 - F(\varsigma_{\text{ABE}}, \tau_{\text{AB}} \otimes \sigma_{\text{E}})^2}$$
$$\leq \sqrt{1 - F(\varsigma_{\text{ABE}}, |\psi\rangle\langle\psi| \otimes \sigma_{\text{E}})^2} \tag{13}$$

$$|\psi\rangle := \sum_{\bm{k}\in K} 2^{-|K|/2}|\bm{k},\bm{k}\rangle \tag{14}$$

(13) is derived by monotonicity of fidelity F under any quantum operations. Next, we consider fidelity where Eve's system is traced out.

$$\sqrt{1-F(\varsigma_{\text{ABE}},|\psi\rangle\langle\psi|\otimes\sigma_{\text{E}})^2}$$
$$\geq \sqrt{1-F(\text{tr}_{\text{E}}\,\varsigma_{\text{ABE}},|\psi\rangle\langle\psi|)^2} \tag{15}$$
$$= \sqrt{1-\langle\psi|(\text{tr}_{\text{E}}\,\varsigma_{\text{ABE}})|\psi\rangle}$$

Compared to (13), (15) has an inequality with opposite direction [16, 17], but [24, 25] suggest to choose $\sigma_{\text{E}}$ so that the equality in (15) holds. Then, entanglement purification is applied to $\text{tr}_{\text{E}}\,\varsigma_{\text{ABE}}$ to obtain a quantum state near (14). Let $\varepsilon_{\text{sec}}^2$ be a failure probability in obtaining such a state,

$$\sqrt{1-\langle\psi|(\text{tr}_{\text{E}}\,\varsigma_{\text{ABE}})|\psi\rangle} \leq \sqrt{1-(1-\varepsilon_{\text{sec}}^2)} = \varepsilon_{\text{sec}} \tag{16}$$

Therefore [24, 25] concludes (7).

### 4.4. Problems with Interpretation of Upper-Bound of Trace Distance

Contrary to the proof in Subsection 4.1, this proof seems to have no unnatural assumptions. However, [22] explained that it is not clarified yet that the security of QKD systems in parallel unless

$$\sigma_{\text{E}} := \sum_{\bm{k}_{\text{A}},\bm{k}_{\text{B}}\in K}\Pr(\bm{k}_{\text{A}},\bm{k}_{\text{B}})\rho_{\text{E}}(\bm{k}_{\text{A}},\bm{k}_{\text{B}}) \tag{17}$$

Actually, M. Tomamichel changed his definition of ε-security [26] from (18) to (19) from what he told [27].

$$\tfrac{1}{2}\min_{\sigma_{\text{E}}}\text{tr}|\rho_{\text{ABE}} - \tau_{\text{AB}} \otimes \sigma_{\text{E}}| \leq \varepsilon_{\text{sec}} \tag{18}$$

$$\tfrac{1}{2}\text{tr}|\rho_{\text{ABE}} - \tau_{\text{AB}} \otimes \sigma_{\text{E}}| \leq \varepsilon_{\text{sec}} \tag{19}$$

If so, we cannot choose $\sigma_{\text{E}}$ to let it hold the equality in (15), which means that $\varepsilon_{\text{sec}}$ cannot be the upper-bound of the trace distance.

Even if we allow to choose $\sigma_{\text{E}}$ so that the equality of (15) holds, it does mean that the trace distance is non-zero. Therefore, $\varsigma_{\text{ABE}} \neq \tau_{\text{AB}} \otimes \sigma_{\text{E}}$, which means that IID of the final key would not be obtained, or means Eve's quantum system is still correlated to the final key, or both.

### 4.5. Interpretation regarding Trace Distance as Indistinguishability of Quantum States

There is an interpretation that the trace distance is "indistinguishability" of the ideal quantum state and the real quantum state [22, 28]. Such an interpretation is justified by citing quantum binary decision theory by C. W. Helstrom [29]. The following is the overview.

In Helstrom's theory, Alice prepare $\rho_0$ or $\rho_1$ with prior probabilities of $p_0$ and $p_1$. Bob discriminates the quantum states with an optimum measurement basis. The maximum guessing probability for Bob is

$$P_{\text{guess}} = \tfrac{1}{2} + \tfrac{1}{2}\text{tr}|p_0\rho_0 - p_1\rho_1| \tag{20}$$

If we assume $p_0 = p_1 = 1/2$, then we see trace distance.

$$P_{\text{guess}} = \tfrac{1}{2} + \tfrac{1}{2}\times\tfrac{1}{2}\text{tr}|\rho_0 - \rho_1| \tag{21}$$

Refs. [28, 22] gave an interpretation as follows. Alice prepares a QKD system which Eve can interact with, or a QKD system with an interface which gives Eve measurement results as if she is interacting with the former system but actually she cannot interact at all. Alice randomly prepares such systems with a prior probability of 1/2, and Eve judges which system is used from her measurement results. If we regard $\rho_0 = \rho_{ABE}$ and $\rho_1 = \tau_{AB} \otimes \sigma_E$, then the maximum guessing probability for Eve is given by (21), therefore the trace distance can be seen as advantage for Eve to distinguish the two situations.

### 4.6. Problems with the Indistinguishability

The problems with the interpretation are as follows. It is said that Eve's success probability in guessing the correct system is given by (21), however, this would not give any idea how high the probability is for Eve to guess the correct key. Furthermore, it gives the following problems.

Now let us think the original situation of QKD. Firstly, $\tau_{AB} \otimes \sigma_E$ is a desirable quantum state but it cannot be distributed, and $\rho_{ABE}$ is the quantum state always distributed. However, in the context of indistinguishability, Alice and Bob have to prepare such quantum states with a prior probability of 1/2. Such a situation does not meet the actual situation of QKD. Furthermore, if they could prepare the system with which Eve cannot interact with a probability of 1/2, then we are not sure why they do not use always, while Eve is sure that the system is always the one which she can interact. This means, the prior probability is very contrived. One more thing, in the quantum binary decision theory, Bob receives the whole system of the quantum state, but in the context of QKD, Eve receives only the partial system of the quantum state, such as $\mathrm{tr}_{AB}\, \rho_{ABE}$. Therefore, the physical situation of the quantum binary decision problem is different from the situation of QKD.

From above, the indistinguishability interpretation cannot be useful to evaluate the security of QKD.

## 5. Security of QKD from Attacker's View
### 5.1. Guessing Probability for Eavesdropper

Ref. [22] derived the following inequality as well, although it did not emphasize its use.

$$\Pr(K|E) \leq 2^{-|K|} + \tfrac{1}{2}\mathrm{tr}|\varsigma_{ABE} - \tau_{AB} \otimes \sigma_E| \leq 2^{-|K|} + \varepsilon_{sec} \quad (22)$$

This is the average probability for Eve to guess the correct shared key from her measurement result $E$. The derivation is as follows. Trace distance generally has the following relation.

$$\mathrm{tr}\left[\Gamma(\varsigma_{ABE} - \tau_{AB} \otimes \sigma_E)\right] \leq \tfrac{1}{2}\mathrm{tr}|\varsigma_{ABE} - \tau_{AB} \otimes \sigma_E| \quad (23)$$

Now, introduce a projection operator

$$\Gamma := \sum_k |k,k\rangle\langle k,k| \otimes M_k \quad (24)$$

Therefore,

$$\begin{aligned}&\mathrm{tr}\left[\Gamma(\varsigma_{ABE} - \tau_{AB} \otimes \sigma_E)\right] \\ &= \sum_{k_A,k_B}\left[\Pr(k_A,k_B)\Pr(k_A|E)\right] - 2^{-|K|} \\ &= \Pr(K|E) - 2^{-|K|}\end{aligned} \quad (25)$$

By transposition of $2^{-|K|}$, we obtain (22). Note that this derivation is independent of specific protocols, implementations, and procedures of security proofs as long as the security measure is the trace distance.

This result clearly shows that the average guessing probability for Eve can be larger than "the maximum failure probability" obtained from the trace distance alone by the constant $2^{-|K|}$. Also this result corresponds to what Yuen pointed out in [7].

### 5.2. Meaning of Attacker's Guessing Probability

In this subsection, the meaning of (22) on the security of QKD is discussed.

Ref. [22] gave an example $\varepsilon/|K| = 10^{-24}$ which would guarantee the security for an age of the universe. Recent experiment [30] claimed that the trace distance of $2^{-50} = 8.9 \times 10^{-16}$ was achieved.

On the other hand, the standard length of the final key is typically about or below $|K| = 10^6$ bits [26]. Then,

$$\Pr(K) = 2^{-|K|} = 2^{-10^6} \ll \Pr(K|E) \sim \varepsilon_{sec} = 2^{-50} \quad (26)$$

The "maximum failure probability" interpretation has been telling us that the probability distribution of the final key is $2^{-|K|}$ with a probability of $1 - \varepsilon_{sec}$. However, Eve's maximum average guessing probability in (22) and its numerical example (26) mean that the obtained probability distribution of the final key for Eve is far from IID. Only when $\varepsilon_{sec} = 0$, we can say the distributed key is perfect. Therefore, the obtained keys from QKD systems do not satisfy the condition of the perfect security $P(K|E) = P(K)$ explained in Section 2. This affects the concept of Universal Composability [3-5, 21, 22, 28] by which we could regard any portion of the final key would be statistically independent form other parts so we could use the portion for any use such as message authentication apart from message encryption, however, like above, IID will not be obtained from Eve's view as long as $\varepsilon_{\text{sec}} \neq 0$.

There may be opinion that we could use such keys if Eve's guessing probability is sufficiently small even if the key is not perfect. Therefore we discuss how secure the key is when $\varepsilon_{\text{sec}} = 2^{-50} = 8.9 \times 10^{-16}$ [30].

Assume that a QKD system is running for 24 hours 365 days, at the communication speed of $10^9$ bits/sec with the final key length $10^6$ bits. Then, $3 \times 10^{10}$ keys will be exchanged in a year. Since $2^{-|K|} \ll \varepsilon_{\text{sec}}$, the expected number of keys leaked to Eve is $3 \times 10^{-5}$. This number looks sufficient.

On the other hand, $7.5 \times 10^3$ traffic fatal accidents had been reported in 2008 in Japan [31], while there were $7.9 \times 10^7$ cars in the same year [32]. Therefore, one car caused $9.5 \times 10^{-5}$ traffic fatal accidents in average in 2008.

The above values show that the number of potential eavesdropping on one QKD system sees in a year is at the same order of magnitude of traffic fatal accidents one car may causes in a year. If QKD systems spread over the world as explained in the introduction, the number of potential eavesdropping is close to the number of traffic fatal accidents.

Even when we obtain $\varepsilon_{\text{sec}} = 2^{-50}$, its security level is still insufficient for widely used important infrastructures.

### 5.3. Other problems

Furthermore, Yuen pointed out many important problems in QKD. For example, in Ref. [33], the trace distance is averaged over exchanged information on the classical channel such as privacy amplification codes, say $y$.

$$\sum_y P_{\text{pub}}(y) \tfrac{1}{2} \text{tr} \left| \rho_{\text{ABE}|y} - \tau_{\text{AB}|y} \otimes \sigma_{\text{E}|y} \right|$$
$$\leq \sum_y P_{\text{pub}}(y) \sqrt{2 P_{\text{ph}|y}} \leq \sqrt{\sum_y P_{\text{pub}}(y) 2 P_{\text{ph}|y}} \quad (27)$$
$$= \sqrt{2 P_{\text{ph}}} =: \varepsilon$$

Actually, however, Eve always knows which codes were used during the post processes to obtain the final key. She is not seeing the averaged information on the classical channel. Therefore we have to evaluate the individual trace distance appeared in the summation, before being averaged. To evaluate the individual trace distance, Yuen proposes to apply Markov inequality multiple times. For instance, if the trace distance is averaged over possible Privacy Amplification Codes, one needs to apply one Markov inequality, resulting in $2\varepsilon_{\text{sec}}^{1/2}$ instead of $\varepsilon_{\text{sec}}$. If Error Correction Codes could be chosen randomly, or if we take Known-Plaintext Attack (KPA) later explained into consideration, one needs to apply Markov inequality twice, resulting in $3\varepsilon_{\text{sec}}^{1/3}$ instead of $\varepsilon_{\text{sec}}$.

Furthermore, KPA [18-20] has to be taken into consideration. KPA is a situation where Eve knows some parts of the plaintext Alice and Bob exchange, therefore she obtains the part of the secret key and tries to guess the rest of the key string. Actually, during the World War II, Japan's cipher called purple cipher was breached because some plaintexts were also sent by red cipher, which was already breached, therefore KPA was possible [34]. A more familiar case is the header of data packets on the internet, which are almost fixed phrases therefore Eve may be able to guess. Remember, that the trace distance was introduced in Ref. [3] to make QKD secure under KPA, because negligible mutual information still allowed Eve to launch KPA to obtain the whole key. Under "the failure probability interpretation", the rest of the key string is independent from the known part, so the keys were assumed to be secure even under KPA [22]. However, now we know this interpretation is incorrect and the bits in the key is not statistically independent. On this problem, Yuen already has given some answers.

Finally, Yuen pointed out that we have to consider the situation where Eve may obtain a key which is close to the correct secret key, while Eve knows how many bit errors are there. If Eve decrypts OTP with this nearly correct key, she sees an almost perfect plaintext with some bit errors. Yuen named this issue "Bit-Error-Rate issue," to mean how many bit errors are allowed for Eve to let her read the plaintext [8]. This problem gives Eve greater possibility compared to the probability to guess the exact key in (22). Even Yuen says how we give an upper-bound on this probability is an open question.

Yuen also pointed out many issues on QKD. The author would ask readers to check other Yuen's papers for more problems in details.

### 6. Conclusions

The trace distance in QKD has been interpreted as a failure probability in distributing ideal keys.

However, there has been criticisms that H. P. Yuen and O. Hirota have repeatedly criticized that such an interpretation is not correct since 2009. The author of this study has been treating this issue since May 2014, QIT30 conference. Now, we obtained the average guessing probability for the eavesdropper on the shared key, therefore we now can conclude the claim Yuen and Hirota have been warning was correct because the success probability of eavesdropping was derived.

There is also indistinguishability interpretation on the trace distance, however, we also see that this interpretation is not useful.

The author would appreciate if these problems could be treated seriously to consider the security level of Quantum Key Distribution.

## 文　献

[1] C. H. Bennett, G. Brassard, "Quantum cryptography: Public key distribution and coin tossing," Proceedings of IEEE International Conference on Computers, Systems and Signal Processing, Vol. 175. No. 0. (1984), or Theoretical Computer Science 560 7-11 (2014).
[2] C. H. Bennett, et al. "Generalized privacy amplification," Information Theory, IEEE Transactions on 41.6 1915-1923 (1995).
[3] R. König, et al. "Small accessible quantum information does not imply security." Physical Review Letters 98.14 140502 (2007).
[4] R. Renner, R. König, "Universally composable privacy amplification against quantum adversaries," Theory of Cryptography. Springer Berlin Heidelberg, 407-425 (2005).
[5] V. Scarani, et al. "The security of practical quantum key distribution," Reviews of modern physics 81.3 1301 (2009).
[6] H. P. Yuen, "Universality and The Criterion 'd' in Quantum Key Generation," arXiv:0907.4694v1, quant-ph (2009).
[7] H. P. Yuen, "Fundamental Quantitative Security In Quantum Key Distribution," Physical Review A 82.6 062304 (2010).
[8] H. P. Yuen, "What The Trace Distance Security Criterion in Quantum Key Distribution Does And Does Not Guarantee," arXiv:1410.6945v1, (2014).
[9] H. P. Yuen, "Problems of Security Proofs and Fundamental Limit on Key Generation Rate in Quantum Key Distribution," arXiv:1205.3820v2, quant-ph (2012).
[10] O. Hirota, "Incompleteness and Limit of Quantum Key Distribution Theory," arXiv:1208.2106v2, quant-ph (2012).
[11] Updating Quantum Cryptography and Communications2015 in QCrypt2015 (2015).
[12] Press-release from Toshiba Corporation, http://www.toshiba.co.jp/about/press/2015_06/pr_j1801.htm (2015).
[13] T. Iwakoshi and O. Hirota, "Problem with Interpretation of Trace Distance as Failure Probability in Quantum Key Distribution (in Japanese)," The 30th Quantum Information Technology Symposium, IEICE QIT2014-9 (2014).
[14] T. Iwakoshi and O. Hirota, "Problem with Interpretation of Trace Distance as Failure Probability in Quantum Key Distribution: Part II (in Japanese)," IEICE ISEC2014-48 (2014).
[15] T. Iwakoshi and O. Hirota, "Misinterpretation of statistical distance in security of quantum key distribution shown by simulation," SPIE Security + Defence. International Society for Optics and Photonics (2014).
[16] T. Iwakoshi, "Defect in Security Proof of Quantum Key Distribution based on Complementarity -Problem with Interpretation of Trace Distance as Failure Probability in QKD: Part III- (in Japanese)," The 31st Quantum Information Technology Symposium, IEICE QIT2014-58 (2014).
[17] T. Iwakoshi, "Defect in Security Proof of Quantum Key Distribution based on Complementarity: Part II (in Japanese)," The 32nd Quantum Information Technology Symposium, IEICE QIT2015-16 (2015).
[18] D. R. Stinson, "Cryptography: Theory and Practice, Third Edition, Edition 3," CRC Press, (2005).
[19] A. J. Menezes, P. C. van Oorschot, and S. A. Vanstone, "Handbook of applied cryptography," CRC Press (1997).
[20] J. A. Buchmann, translated by Y. Hayashi, "Introduction to Cryptology (in Japanese)," 3rd Edition, Springer Japan, (2008).
[21] R. Renner, "Security of Quantum Key Distribution," International Journal of Quantum Information 6.01, 1-127 (2008).
[22] C. Portmann, R. Renner, "Cryptographic security of quantum key distribution," arXiv:1409.3525v1, quant-ph (2014).
[23] K. Kato, "Coupling Lemma and Its Application to The Security Analysis of Quantum Key Distribution," arXiv:1505.06269v1, quant-ph (2015), or Tamagawa University Quantum ICT Research Institute Bulletin, vol.4, no.1, pp.23-30, (2015).
[24] M. Koashi, "Simple security proof of quantum key distribution based on complementarity," New Journal of Physics 11.4 045018 (2009).
[25] M. Koashi and T. Koshiba, "Proceeds in Quantum Cryptography (in Japanese)," Saiensu-sha Co., Ltd. Publishers, ISSN0386-8257, (2008).
[26] M. Tomamichel, et al., "Tight finite-key analysis for quantum cryptography," Nature communications 3 634 (2012).
[27] M. Tomamichel, et al., "Tight finite-key analysis for quantum cryptography," arXiv:1103.4130v2, quant-ph (2012).
[28] M. Ben-Or, et al. "The universal composable security of quantum key distribution." Theory of Cryptography, Springer Berlin Heidelberg, p.386-406. (2005).
[29] C. W. Helstrom, "Quantum Detection and Estimation Theory," Journal of Statistical Physics 1.2 231-252 (1969) or Academic press (1976).
[30] H. Takesue, et al., "Experimental quantum key distribution without monitoring signal disturbance," Nature Photonics (2015).

[31] Ministry of Health, Labour and Welfare Japan
http://www.mhlw.go.jp/toukei/saikin/hw/jinkou/tokusyu/furyo10/01.html

[32] Automobile Inspection and Registration Information Association Japan
https://www.airia.or.jp/publish/file/e49tph00000004sb-att/e49tph00000004si.pdf

[33] M. Hayashi, T. Tsurumaru, "Concise and tight security analysis of the Bennett–Brassard 1984 protocol with finite key lengths," New Journal of Physics 14.9 093014 (2012).

[34] T. Suita, "All about cryptography (in Japanese)," Gijutsu-Hyohron Co., Ltd., ISBN4-7741-0616-X, (1998)